\documentstyle[12pt]{article}
\input{psfig}
\newcommand{\pom}{I\!\! P}
\begin{document}
\begin{center}
{\large\bf Diffraction in QCD}
\vglue 0.3cm
{K.  GOULIANOS\\ }
\baselineskip=13pt
{\em The Rockefeller University, 1230 York Avenue\\
 New York, NY 10021, USA\\
{\small \vglue 1em
Presented at CORFU-2001, 
Corfu, Greece, 31 Aug - 20 Sept 2001}\\
}
\end{center}

\centerline{\bf Abstract}

{\rightskip=10pc
\leftskip=10pc
Results on soft and hard diffraction
are briefly reviewed and placed in a QCD perspective using a parton 
model approach. Issues addressed include factorization, scaling properties,
universality of rapidity gap formation, and unitarity.
Predictions for differential cross sections of processes 
with multiple rapidity gaps are presented with examples 
for the Tevatron and the Large Hadron Collider.
This paper is an expanded version of a paper delivered at 
``Snowmass-2001"~\cite{snowmass}.  
}
\section{Introduction}
Experiments at $\bar p(p)$-$p$  and $e$-$p$ colliders have
reported and characterized a class of events incorporating a hard
(high transverse momentum) partonic scattering while carrying the
characteristic
signature of diffraction, namely a leading (anti)proton and/or
a large ``rapidity gap", defined as a region of pseudorapidity,
$\eta\equiv -\ln(\mbox{tan}\frac{\theta}{2})$,
devoid of particles (see Figs.~1 and 2). 
The rapidity gap is a non-perturbative phenomenon presumed to be caused by the
exchange of a Pomeron~\cite{Regge}, whose generic QCD definition is
a color-singlet combination of
quarks and/or gluons carrying the quantum numbers of the vacuum.
\vglue -0.7in
\centerline{
\hspace*{0.75cm}\psfig{figure=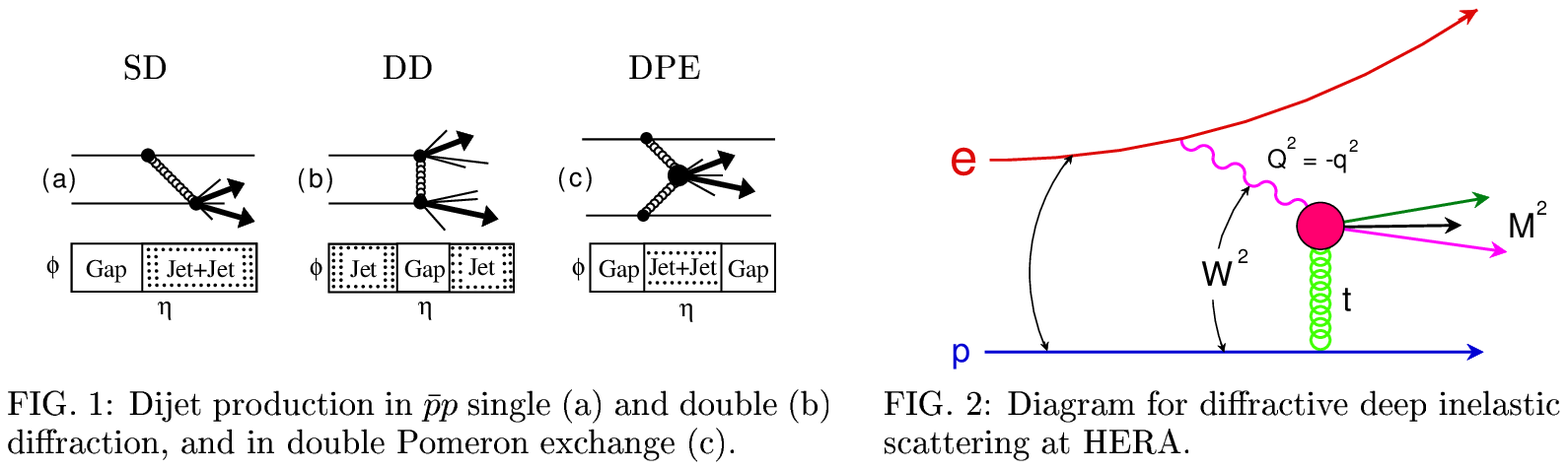,width=8in}}
\vglue -7.3in
The interplay between soft and hard processes in {\em hard diffraction}
is of particular theoretical importance due to its
potential for elucidating
the transition from perturbative to nonperturbative QCD. 
Phenomenological models proposed for hard diffraction 
have been only partially successful in describing the data. 
A QCD-based theoretical description is still not available. 
This is not surprising, since diffraction
invariably involves non-perturbative effects associated with the
formation of rapidity gaps. 
The theoretical community has thus far paid very little attention 
to {\em soft} diffractive processes, i.e. those which do not have a hard
partonic scattering. Yet, experiment has shown that soft processes 
have ``a lot to say" about rapidity gap formation, which is key to 
understanding diffraction.
In this paper, we examine the regularities observed in both soft and 
hard diffractive processes using a parton model approach, 
which places diffraction in a QCD perspective.

\section{Rapidity gaps}
The exchange of a gluon or a quark between two hadrons
at high energies leads to events in which, in addition to whatever
hard scattering may have occurred, the entire rapidity space is
filled with soft (low transverse momentum) particles
{\em (underlying event)}.
The soft particle distribution is approximately flat in
(pseudo)rapidity.
The flat $dN/d\eta$ shape is the result of $x$-scaling~\cite{Feynman} of
the parton distribution functions of the incoming hadrons.
Rapidity gaps can be formed in any inelastic
non-diffractive (ND) event by multiplicity fluctuations.
Assuming Poisson statistics, the probability for a gap of
width $\Delta\eta$ within a ND event sample is given by
\begin{equation}
P^{ND}_{gap}(\Delta \eta)=\rho\,e^{-\rho\,\Delta \eta}
\label{NDG}
\end{equation}
\noindent where $\rho$ is the average particle density per unit $\eta$
(the total probability for {\em any} gap is normalized to unity).
As seen in~(\ref{NDG}), the probability for 
rapidity gaps formed my multiplicity fluctuations is
exponentially suppressed. In contrast, since no radiation is emitted by 
the acceleration of vacuum quantum numbers,
Pomeron exchange leads to large rapidity gaps whose probability is not
exponentially dumped. Therefore, large rapidity gaps
are an unmistakable diffractive signature and 
can be considered the {\em generic} definition of diffraction.

\section{Soft diffraction}
Soft diffraction has been traditionally treated theoretically in the framework 
of Regge theory. For large rapidity gaps ($\Delta \eta\geq 3$), 
the cross sections for $\bar p(p)$-$p$ single and double (central) diffraction  
can be written as~\cite{dd}
\begin{equation}
{d^{2}\sigma_{SD}\over dtd\Delta \eta}=
  \left[{\beta^{2}(t)\over 16\pi} e^{2[\alpha(t)-1]\Delta \eta}\right]
\!\left[\kappa\beta^{2}(0)
{\left(\frac{s'}{\textstyle s_{\circ}}\right)}^{\epsilon}\right],
\;\;{d^{3}\sigma_{DD}\over dtd\Delta \eta d\eta_c}=
\left[{\kappa\beta^{2}(0)\over 16\pi} e^{2[\alpha(t)-1]\Delta \eta}\right]
\!\left[\kappa\beta^{2}(0){{\left(\frac{s'}{\textstyle s_{\circ}}\right)}}^
{\epsilon}
\right]\\
\label{sddd}
\end{equation}
\noindent where $\alpha(t)=1+\epsilon+\alpha't$
is the Pomeron trajectory, $\beta(t)$ the coupling of the
Pomeron to the (anti)proton, and $\kappa\equiv g(t)/\beta(0)$ the ratio
of the triple-Pomeron to the Pomeron-proton
couplings. 
The above two equations, which are based on factorization, 
are remarkably similar. In each case, 
there are two factors:
\begin{itemize}
\item the first, which is 
$\sim \{{\rm exp}[(\epsilon+\alpha't)\Delta\eta]\}^2$ and thus depends on 
the rapidity gap,
\item the second, which has the form $\sim (s')^{\epsilon}$ 
of a total cross section at c.m.s. energy squared $s'$.
\end{itemize}
In single diffraction (SD), 
the second factor is interpreted as the Pomeron-nucleon 
total cross section, while the first factor as the square of the elastic
scattering amplitude between the diffractively excited nucleon state and the 
other nucleon. The similarity between the SD and double diffraction (DD) 
equations suggests 
an interpretation of the first factor in DD as the elastic scattering 
between the two diffractively dissociated nucleon states. However, 
the interpretation of the second factor in DD is not as straightforward 
as in SD, but does acquire a similar meaning in the parton model, as 
discussed in the next section. 
Since the second factors in~(\ref{sddd}) represent properly
normalized cross sections, the first factors may be thought of 
as the rapidity gap probability and {\em renormalized} to 
unity. A model based on such a renormalization 
procedure~\cite{RR,KGgap} has yielded 
predictions in excellent agreement with measured SD and DD cross 
sections, as seen in Figs.~3 and 4. 

\vglue -0.7in
\centerline{\psfig{figure=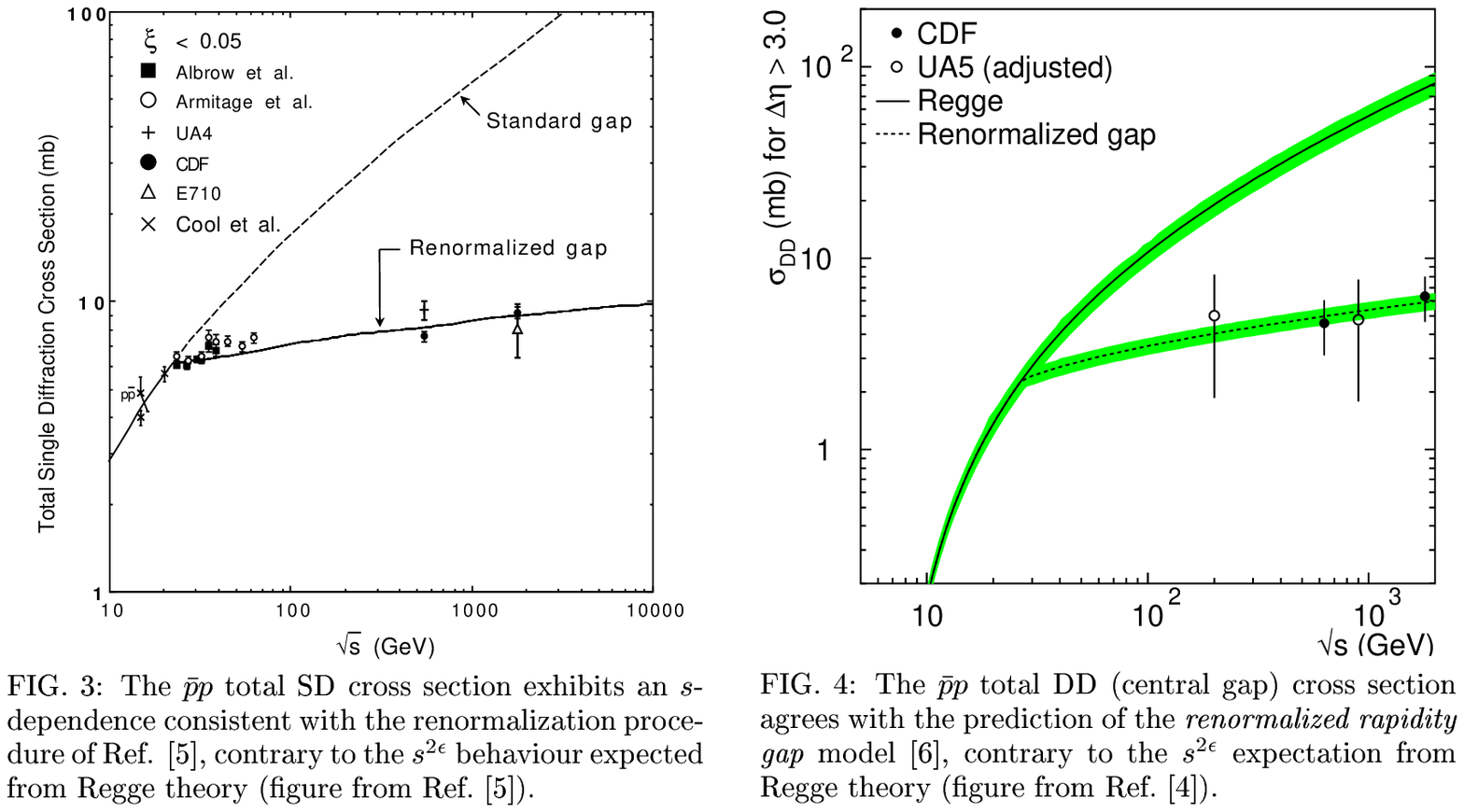,width=8in}}
\vspace*{-6in}

\section{Factorization, scaling and unitarity}

The renormalized rapidity gap
probability is by definition energy independent and thus represents 
a scaling behaviour, which is most prominently displayed in the form of 
the SD differential cross section in terms of the square of the mass of the 
diffractively excited nucleon state, $M^2$. The variable $M^2$ is related to 
$\Delta\eta$ and to the fractional momentum loss of the leading nucleon: 
$M^2=s\xi=se^{-\Delta\eta}$. The unrenormalized SD cross section 
at $t=0$ has the form 
$\frac{d\sigma_{SD}}{dM^2}|_{t=0}
\propto\frac{s^{2\epsilon}}{(M^2)^{1+\epsilon}}$. 
This form 
leads to a total diffractive cross section $\sim s^{2\epsilon}$, which grows 
faster than the total nucleon-nucleon cross section ($\sim s^{\epsilon}$) and 
would exceed it at $\sqrt s\approx $2 TeV  violating unitarity 
(see Figs.~3 and 4). The renormalization procedure leads to the 
energy independent form 
$\frac{d\sigma_{SD}}{dM^2}|_{t=0}\propto\frac{1}{(M^2)^{1+\epsilon}}$,
which avoids the violation of unitarity. This form is brilliantly 
supported by the data, as can be seen in Fig.~5.  
Thus, it appears that nature preserves unitarity by favoring 
``$M^2$-scaling" over factorization.

\centerline{\psfig{figure=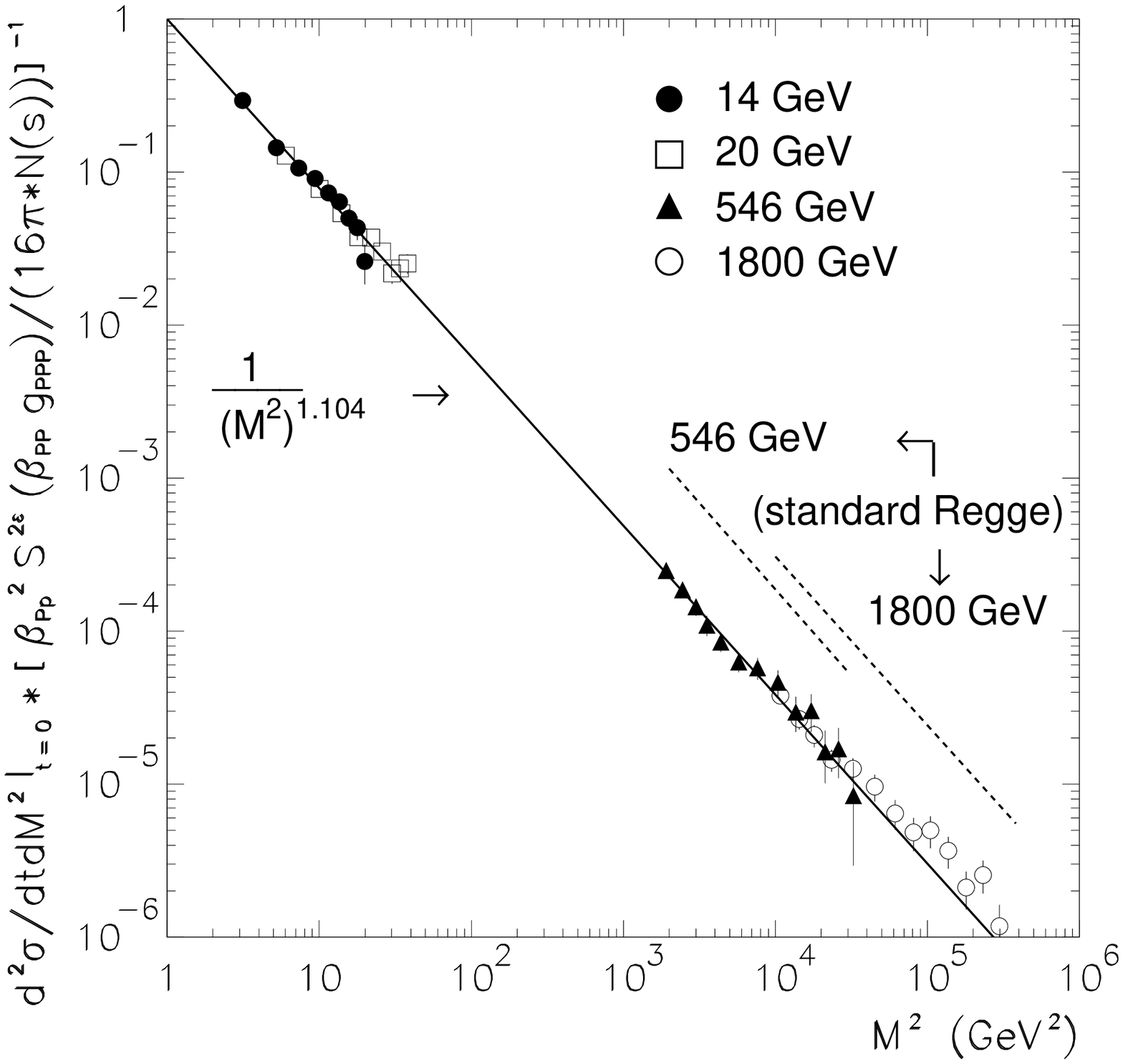,width=4in}}
Fig.~5: {Cross sections \protect$d^2\sigma_{SD}/dM^2 dt$
for $p+p(\bar p) \rightarrow p(\bar p)+X$ at
$t=0$ and $\protect\sqrt s=14$, 20, 546 and 1800 GeV, multiplied by
$[\beta^2_{\pom pp}\,s^{2\epsilon}(\beta_{\pom pp}\cdot
g_{\pom\pom\pom})/(16\pi *N(s))]^{-1}$, where $N(s)$
is the integral of the rapidity gap probability,
are compared with the renormalized gap
prediction of \protect$1/(M^2)^{1+\epsilon}$ using
$\epsilon=0.104$~\protect\cite{CMG}. The dashed curves show the
standard Regge-theory predictions.
The $t=0$ data were obtained by
extrapolation from their $t=-0.05$ GeV$^2$ values after subtracting the
pion exchange contribution~\protect\cite{GM}.}
\vglue -2in
\section{A parton model approach}
In the parton model, the $pp$ total cross section
is basically proportional to the number of partons in the proton,
integrated down to $x=s_{\circ}/s$, where $s_{\circ}$ is the
energy scale for soft physics.
The latter is of ${\cal{O}}(\langle M_T\rangle^2)$,
where $\langle M_T\rangle\sim 1$ GeV
is the average transverse mass of the particles in the final state.
Expressing the parton density as a power law in $1/x$, which is an appropriate
parameterization for
the small $x$ region responsible for the cross section rise at
high energies, we obtain
\begin{equation}
\sigma_T\sim
{\displaystyle\int}_{(s_{\circ}/s)}^1\;\frac{dx}{x^{1+n}}
\sim\left(\frac{s}{s_{\circ}}\right)^{n}
\end{equation}
The parameter $n$ is identified as the
$\epsilon=\alpha(0)-1$ of the Pomeron trajectory~\cite{Levin}. 

According to the optical theorem, the form $s^{\epsilon}$ of the 
total cross section is that of the imaginary part of the forward ($t=0$) 
elastic scattering amplitude.  Note that $\ln s$ is the (pseudo)rapidity 
region $\Delta\eta$ in which there is particle production. The full parton 
model amplitude can be written as 
\begin{equation}
{\rm Im\,f}(t,\Delta y)\sim e^{(\textstyle{\epsilon}+\alpha't)\Delta y}
\label{pma}
\end{equation}
\noindent where we have added to $\epsilon$ the term  $\alpha't$ as 
a parameterization of the $t$-dependence of the amplitude. 

One may now understand expressions (\ref{sddd}) for the SD 
and DD cross sections as follows:
the second term in brackets represents the nucleon-nucleon 
total cross section at a sub-energy squared $s'$, multiplied by a factor 
$\kappa$, which may be interpreted as ``the price to pay" for the 
color matching required to enable the formation of a rapidity gap, while
the first term represents the amplitude squared of the elastic scattering 
between the diffractively dissociated nucleon and the other 
nucleon in SD, or between the two diffractively dissociated nucleons in DD.
Since the cross section represented by the second term is properly 
normalized, the first term simply represents 
a rapidity gap probability distribution 
and should be normalized to unity. This result is equivalent to 
renormalizing the rapidity gap probability of the Regge model,
as proposed in~\cite{RR,KGgap}.

\section{Multiple rapidity gaps in diffraction}
The parton model approach used above to calculate the SD and DD 
differential cross sections lends to easy generalization to events 
with multiple rapidity gaps. Here, we outline the 
procedure for calculating the differential cross section for a 4-gap 
event (see Fig.~6).  

\vglue -1in
\centerline{\psfig{figure=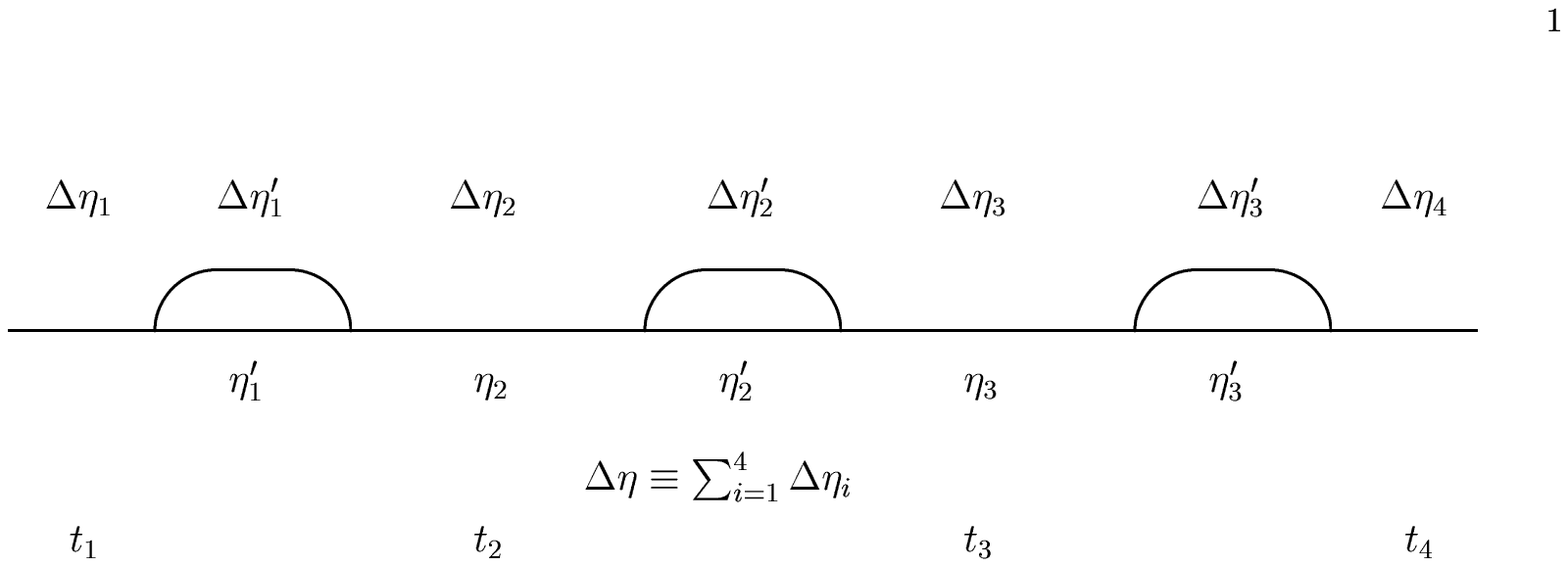,width=8in}}
\vspace*{-7.75in}

\centerline{Fig.~6: Topology of a 4-gap event in pseudorapidity space.}
\vglue 1em
The calculation of the differential cross section 
is based on the parton-model scattering amplitude of Eq.~(\ref{pma}).
For the rapidity regions $\Delta \eta'_i$, where there is  
particle production, the $t=0$ parton model amplitude is used and the 
{\em sub-energy cross section} is  
$\sim e^{\textstyle{\epsilon}\Delta \eta'}$. 
For rapidity gaps $\Delta \eta$ which can be considered as resulting from 
elastic scattering between diffractively excited states, 
the square of the full
parton-model amplitude is used, 
$e^{2(\textstyle{\epsilon}+\alpha't_i)\Delta \eta_i}$, and the form factor 
$\beta^2(t)$ is included for a surviving nucleon.
The {\em gap probability} (product of all rapidity gap terms) 
is then normalized to unity, and 
a {\em color matching factor} $\kappa$ is included for each gap.

For the 4-gap example of Fig.~6, 
which has 10 independent variables 
$V_i$ (shown below the figure), we have:
\begin{itemize}
\item 
$\frac{d^{10}\sigma}{\Pi_{i=1}^{10}dV_i}=
P_{gap}\times \sigma({\rm sub-energy})$
\item 
$\sigma({\rm sub-energy})=
\kappa^4\left[\beta^2(0)\cdot e^{\textstyle{\epsilon}
\Delta y'}\right]$
\hfill ($\Delta y'=\sum_{i=1}^3\Delta \eta'_i$)
\item 
$P_{gap}=N_{gap}\times \Pi_{i=1}^4\left[e^{(\epsilon+\alpha't_i)
\Delta \eta_i}\right]^2\times [\beta(t_1)\beta(t_4)]^2=N_{gap}
\cdot e^{2\textstyle{\epsilon}\Delta \eta}\cdot f(V_i)|_{i=1}^{10}$
\hfill ($\Delta \eta=\sum_{i=1}^4\Delta \eta_i$)\\
where $N_{gap}$ is the factor that normalizes $P_{gap}$
over all phase space to unity.
\end{itemize}
The last equation shows that the renormalization 
factor of the gap probability 
depends only on $s$ (since $\Delta \eta_{max}=\ln s$) 
and not on the number of diffractive gaps. Thus, the ratio of 
two-gap to one-gap cross sections is expected to be $\approx \kappa$, 
with no additional energy dependent suppression for the second gap. Below, 
we discuss two specific two-gap processes which can be studied at the 
Tevatron. Studies of processes with more than two gaps will have 
to await the commissioning of the LHC.  

\subsection{Double Pomeron exchange (DPE)}
The double Pomeron exchange process is shown below for 
$\bar pp$ collisions (we use rapidity $y$ and pseudorapidity $\eta$
interchangeably):\\

\vglue -1.5in
\centerline{\psfig{figure=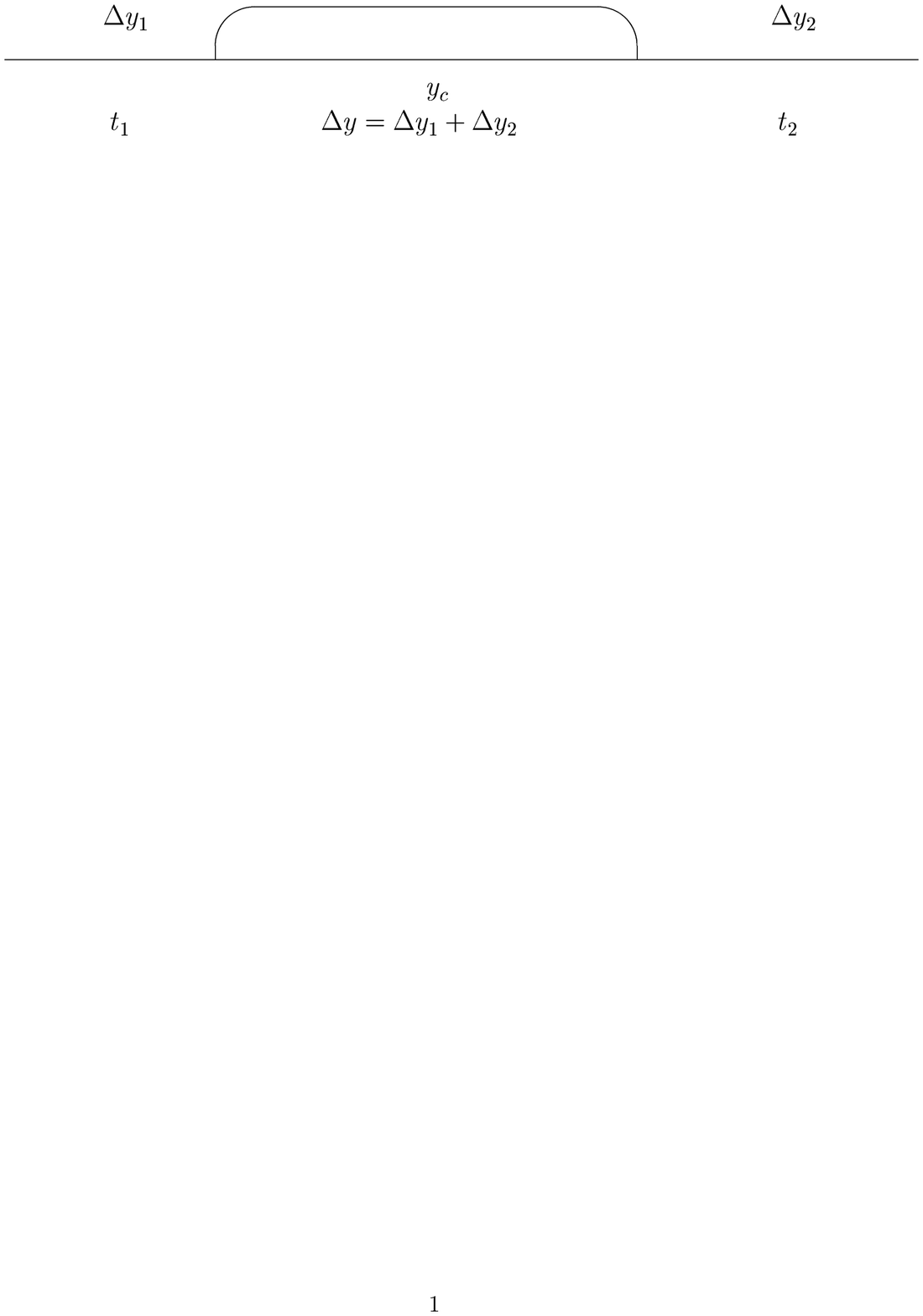,width=8in}}
\vspace*{-7.75in}

\centerline{Fig.~7: 
$\bar p+p\rightarrow \bar p+{\rm GAP}+X+{\rm GAP}+p$}
\vglue 0.15in

The Regge theory predictions for SD and DPE are:\\

\begin{center}
\begin{tabular}{lll}
SD:&
${d^{2}\sigma\over dt\,d\Delta y}=$&
$\;\,\left[{\beta(t)\over 4\sqrt{\pi}}\, e^{[\alpha(t)-1]\Delta y}\right]^2
 \kappa \left\{\beta^2(0)
{\left(\frac{\textstyle{s'}}
{\textstyle s_{\circ}}\right)}^{\alpha(0)-1}\right\}$\\
&&\\
DPE:&
${d^{4}\sigma\over dt_1dt_2d\Delta y dy_c}=$
&
$\Pi_i\left[{\beta(t_i)\over 4\sqrt{\pi}}
e^{[\alpha(t_i)-1]\Delta y_i}\right]^2$
$\kappa^2\left[\beta^{2}(0){{\left(\frac{\textstyle{s'}}
{\textstyle s_{\circ}}\right)}}^
{\alpha(0)-1}\right]$\\
&Note:&$\Delta y_1=\frac12(\Delta y+y_c);\;\Delta y_2=\frac12(\Delta y-y_c)$\\
&&\\
\end{tabular}
\end{center}

Following the procedure outlined above, the DPE differential cross section can 
be written as\\

\begin{center}
\begin{tabular}{ll}
${d^{4}\sigma\over dt_1\,dt_2\,d\Delta y\,dy_c}=$&
$P_{gap}(t_1,t_2,\Delta y,y_c)\;\;\times\;\;
\kappa^2 \sigma_{tot}(s')$\\
&\\
$P_{gap}(t_1,t_2,\Delta y,y_c)=$&
${\beta(t_1)\over 4\sqrt{\pi}}
e^{\frac12 [\alpha(t_1)-1] (\Delta y+y_c)}
\times
{\beta(t_i)\over 4\sqrt{\pi}}
e^{\frac12 [\alpha(t_i)-1] (\Delta y+y_c)}$\\
&\\
$\sigma_{tot}(s')=$&
$\left[\beta^{2}(0){{\left(\frac{\textstyle{s'}}
{\textstyle s_{\circ}}\right)}}^
{\alpha(0)-1}\right]\;$, where $\ln s'=\ln s-\Delta y$\\
&\\
$R^{DPE}_{SD}|_{{\rm fixed}\;\xi(\bar p)}$=&
$\frac{N_{DPE}}{N_{SD}}\int_{t_p=0}^{\infty}
\int_{\xi_p=M_0^2/(s\times \xi_{\bar p})}^{0.02}\frac{\beta^2(t_p)}{16\pi}
\,\frac{\kappa }{\xi^{\alpha(0)+2\alpha't_p}}dt_pd\xi_p$\\
\end{tabular}
\end{center}

The renormalization factor $N_{DPE}$ is obtained by integrating the 
gap probability $P_{gap}$ over $t_1,t_2,\Delta y, y_c$. 
The limits of integration are:\\
$0<-t_1<\infty$; $0<-t_2<\infty$\\
$2.3<\Delta y<\ln (s/s_0)$,
where $s_0=1$ GeV$^2$ and $\Delta y=2.3$ corresponds to $\xi=0.1$\\
$-\frac12 (\Delta y-2.3)<y_c<\frac12 (\Delta y-2.3)$\\
For SD, the gap probability is the first term in the SD equation and 
$N_{SD}$ is obtained by integrating it over $\Delta y$  within 
the region $2.3<\Delta y<\ln (s/s_0)$.

For numerical evaluations we use~\cite{CMG,GM} $\alpha(t)=1.104+0.25t$,
$\beta(0)=4.1$~mb$^{\frac{1}{2}}$ (6.57 GeV$^{-1}$) and $\kappa=0.17$.
The ratio of DPE to SD rates in $\bar pp$
collisions with a leading final state antiproton of fractional momentum loss
$\xi_{\bar p}=0.065$, the average value of the
inclusive $\bar p$-triggered data samples in the CDF publications
of Refs.~\cite{jj1800,jj630}, and for $\xi_p>0.02$ is predicted to be
\begin{center}
\fbox{{$R^{DPE}_{SD}|_{\xi_{\bar p}=0.065,\,\xi_p>0.02}
=0.21\;(0.17)\;\pm 10\%$ at 1800 (630) GeV}}
\end{center}
where the error is due the the uncertainty on the factor $\kappa$.
This ratio does not depend strongly on $\xi(\bar p)$ within the range
$0.035<\xi(\bar p)<0.95$ of the CDF event samples and therefore represents a
realistic prediction.

\subsection{Single diffraction with a second gap (SDD)}
An interesting two-gap process is $\bar pp$ single diffraction with 
a leading final state antiproton and a rapidity gap within the 
rapidity space allocated to the diffraction dissociation products of 
the proton (see Fig.~7).

\vglue -1.35in
\centerline{\psfig{figure=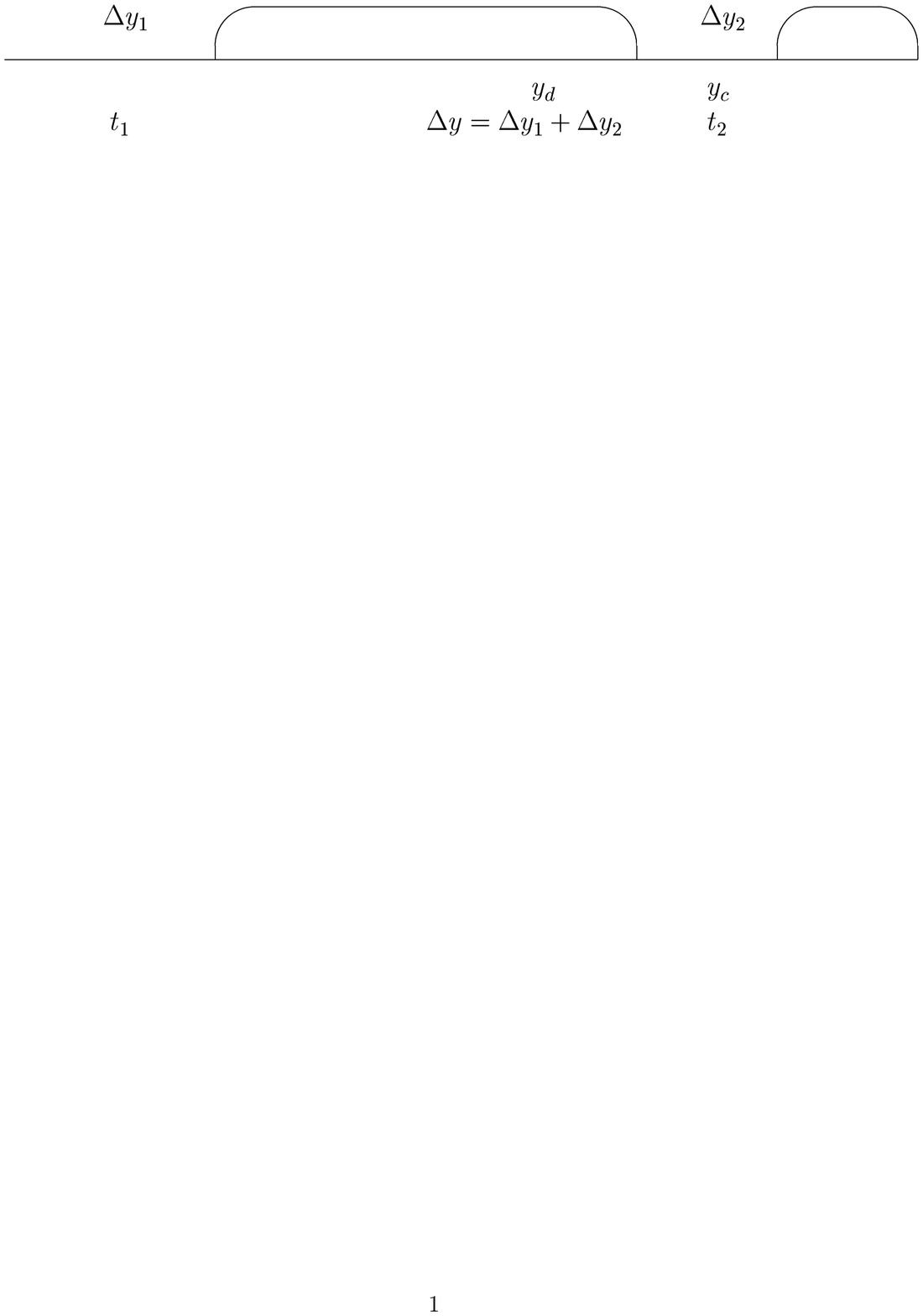,width=8in}}
\vspace*{-7.75in}

\centerline{FIG.~8:   
$\bar p+p\rightarrow \bar p+{\rm GAP}_1+X+{\rm GAP}_2+Y$}
\vglue 0.15in

The second gap, GAP$_2$, may be thought of as due to $\pom$-$p$ 
double diffraction. Thus, this process is a combination of 
single plus double diffraction and hence we represent it by ``SDD".

To calculate the SDD cross section we start with the Regge predictions
for SD, DD and SDD:

\begin{center}
\begin{tabular}{lll}
SD:&
${d^{2}\sigma\over dt\,d\Delta y}=$&
$\;\,\left[{\beta(t)\over 4\sqrt{\pi}}\, e^{[\alpha(t)-1]\Delta y}\right]^2
 \kappa \left\{\beta^2(0)
{\left(\frac{\textstyle{s'}}
{\textstyle s_{\circ}}\right)}^{\alpha(0)-1}\right\}$\\
&&\\
DD:&
${d^{3}\sigma\over dt\,d\Delta y\,dy_c}=$&
$\kappa\left[{\beta(0)\over 4\sqrt{\pi}} e^{[\alpha(t)-1]\Delta y}\right]^2
\kappa\left[\beta^{2}(0){{\left(\frac{\textstyle{s'}}
{\textstyle s_{\circ}}\right)}}^
{\alpha(0)-1}\right]$\\
&&\\
SDD:&
${d^{5}\sigma\over dt_1\,dt_2\,d\Delta y_1\,d\Delta y_2\,dy_c}=$&
$\left[{\beta(t)\over 4\sqrt{\pi}}\, e^{[\alpha(t_1)-1]\Delta y_1}\right]^2
\kappa \left\{
\kappa\left[{\beta(0)\over 4\sqrt{\pi}} e^{[\alpha(t_2)-1]\Delta y_2}\right]^2
\;\kappa\left[\beta^{2}(0){{\left(\frac{\textstyle{s''}}
{\textstyle s_{\circ}}\right)}}^{\alpha(0)-1}\right]\right\}$\\
\end{tabular}
\end{center} 
Following the rules of the parton model approach, the SDD cross section 
may be written as
\begin{center}
\begin{tabular}{ll}
${d^{5}\sigma\over dt_1\,dt_2\,d\Delta y_1\,d\Delta y_2\,dy_c}=$&
$P_{gap}(t_1,t_2,\Delta y_1,\Delta y_2,y_c)\;\;\times\;\;
\kappa^2 \sigma_{tot}(s'')$\\
&\\
$P_{gap}(t_1,t_2,\Delta y_1,\Delta y_2,y_c)=$&
$\left\{
\left[{\beta(t)\over 4\sqrt{\pi}}\,
e^{[\alpha(t_1)-1]\Delta y_1}\right]^2\;
\kappa\left[{\beta(0)\over 4\sqrt{\pi}} e^{[\alpha(t_2)-1]\Delta y_2}\right]^2
\right\}$\\
&\\
$\sigma_{tot}(s'')=$&
$\left[\beta^{2}(0){{\left(\frac{\textstyle{s''}}
{\textstyle s_{\circ}}\right)}}^
{\alpha(0)-1}\right]\;$, where $\ln s''=\ln s-\Delta y_1-\Delta y_2$\\
\end{tabular}
\end{center}
Changing variables, $\Delta y_1,\;\Delta y_2\;\Rightarrow\;
\Delta y=\Delta y_1+\Delta y_2,\;y_d$, 
where $y_d=\frac12\Delta y_1$ is
the center of the single diffractive cluster, the gap probability 
becomes
$$P_{gap}(t_1,t_2,\Delta y,y_d,y_c)=2\kappa
\left(\frac{\beta^2(0)}{16\pi}\right)^2\;e^{2\epsilon\Delta y}
\;F^2_1(t_1)\;e^{2\alpha'(2y_d)t_1}\;e^{2\alpha'(\Delta y-2y_d)t_2}$$
where $F_1(t_1)$ is the antiproton form factor~\cite{RR}. Integrating 
$P_{gap}$ over $t_1,t_2,y_c,y_d,\Delta y$ yields the 
renormalization factor $N_{SDD}$. The limits of integration are:
$0<-t_1<\infty$;
$e^{-(\Delta y-2y_d)}<-t_2<e^{\Delta y-2y_d}$; 
$-\ln\sqrt s+2y_d+\frac12(\Delta y-2y_d)<y_c<
\ln\sqrt s-\frac12 (\Delta y-2y_d)$;
$0<y_d<\frac12 \Delta y$;
$2.3<\Delta y<\ln \frac{\textstyle{s}\,\textstyle{s}_0}{(1.5\,{\rm GeV})^2}$,
where $s_0=1$ GeV$^2$.

Experimentally, in order to be able to detect a large rapidity
gap within a single-diffractive cluster of particles, the cluster must
extend over a large part of the rapidity space covered by the detector.
Therefore, we evaluate the SDD to SD fraction for events with a
relatively large $\xi(\bar p)$-value. For
$\xi_{\bar p}=0.075$, which for $\sqrt s=$1800 (630) GeV corresponds to
sub-energy $\sqrt{s'}\approx 500$ (170) GeV,
the predicted SDD fraction for $\Delta y_2>3$ is
\begin{center}
\fbox{{$R^{two-gap}_{one-gap}|_{\xi_{\bar p}=0.075,\,\Delta y_2>3}=
0.26\;(0.20)\;\pm 10\%$ at 1800 (630) GeV}}
\end{center}
where the error is due the the uncertainty on $\kappa$.
This prediction can be compared with experiment using the existing CDF 
inclusive SD data of  Refs.~\cite{jj1800,jj630}.

\section{Hard diffraction}
The central issue in hard diffraction is the question of QCD factorization:
can hard diffractive cross sections be obtained as a convolution of 
``diffractive structure functions" (DSF) with parton-level cross sections?
This question was addressed decisively by a comparison~\cite{jj1800}
between the DSF measured by CDF in dijet production
at the Tevatron and the prediction based on parton densities extracted
from diffractive DIS at HERA. The DSF at the
Tevatron was found to be suppressed relative to the prediction from HERA by a
factor of $\sim 10$ (see Fig.~9).
This result confirmed previous CDF results from diffractive
$W$~\cite{W}, dijet~\cite{jj} and $b$-quark~\cite{b} production at
$\sqrt s$=1800 GeV, and was corroborated by more recent CDF results
on diffractive $J/\psi$~\cite{jpsi} production at 1800 GeV and
dijet production at 630 GeV~\cite{jj630}.
\vglue -0.05in
\centerline{\psfig{figure=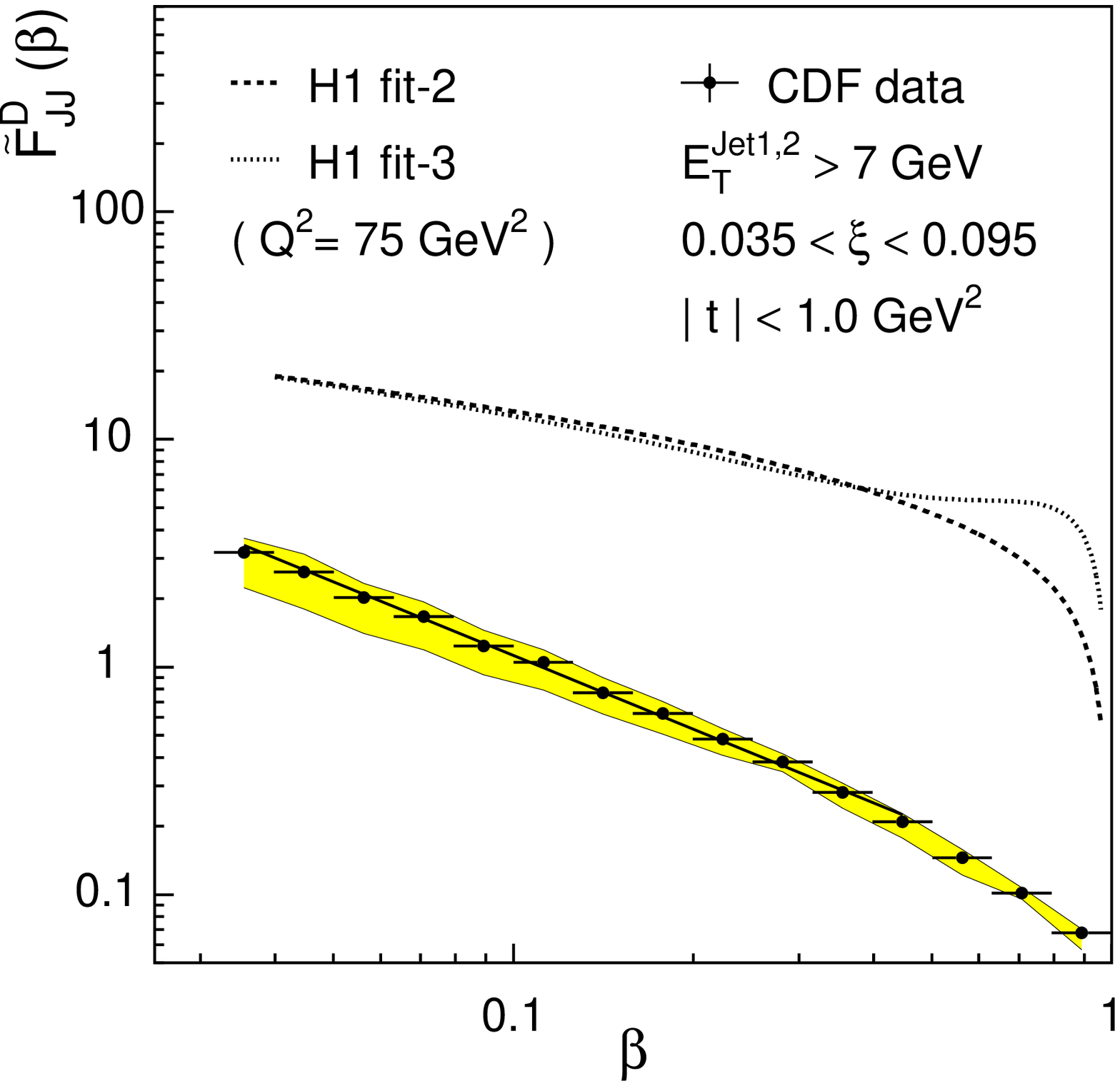,width=4in}}
\vglue -0.5in
\begin{center}
Fig.~9: The diffractive structure function $F_{jj}(\beta,Q^2)$, 
where $\beta$ is the momentum fraction of the parton in the Pomeron, 
extracted from CDF diffractive dijet production data in $\bar pp$ 
collisions at $\protect\sqrt s=$1800 GeV at the Fermilab Tevatron, 
is compared with expectations from
parton densities extracted from diffractive deep
inelastic scattering by the H1 Collaboration at the DESY $ep$ collider HERA
(figure from Ref.~\protect\cite{jj1800}).
\end{center}
        Although factorization breaks down severely between HERA and
the Tevatron, it nevertheless holds within the HERA data and within the
single-diffractive data at the Tevatron at the same center of mass
collision energy. This is demonstrated by the fact that the gluon parton
distribution function (PDF) derived from DIS adequately describes
diffractive dijet production at HERA~\cite{jjhera},
while at the Tevatron a consistent gluon PDF
is obtained from the measured rates of diffractive $W$, dijet,
$b$-quark and $J/\psi$ production~\cite{jpsi}.

Factorization was also tested at the Tevatron
between the structure functions measured in single-diffractive and
double-Pomeron exchange (DPE) dijet production 
at $\sqrt s=$1800 GeV~\cite{DPE}. The ratio of the DPE to SD structure 
functions was found to be larger than that of the SD to ND ones by 
a factor of $5.3\pm 2.0$. This result represents a breakdown of 
factorization within Tevatron data. However, we note that DPE is a two-gap 
process and, according to the parton model approach to diffraction that 
we have presented, should not be suppressed (except for 
kinematical edge effects) relative to SD.
The above result, within the experimental uncertainty, 
confirms the prediction of the parton model approach. 

Finally, CDF has reported that the $\beta$ and $\xi$ dependence of the
diffractive structure function at $\sqrt s=1800$ GeV in the region 
$0.035<\xi<0.095$ obeys $\beta$-$\xi$ factorization for $\beta<0.5$
The observed $\sim\xi^{-1}$ dependence shows
that Pomeron-like behaviour extends to
moderately high $\xi$ values in diffractive dijet production.
Such behaviour is expected in the parton model approach 
for hard diffraction, where
the Pomeron emerges from the quark-gluon sea as a combination of two
partonic exchanges, one on a hard scale that produces the dijet system
and the other on a soft scale that neutralizes the color flow and forms
the rapidity gap~\cite{newapproach}.

\section{Conclusion}
Soft hadronic processes have traditionally been treated 
theoretically in the framework of Regge theory. By postulating a 
Pomeron with a trajectory of intercept $\alpha(0)>1$, 
Regge theory correctly predicts several features of soft physics including 
the rise of total cross sections with increasing energy,
the shrinking of the forward elastic scattering peak, 
elastic to total cross section ratios, and the shape of the
single diffraction differential cross section.
However, the predicted $\sim s^{2\epsilon}$ rise of the 
single diffractive  cross section is 
faster than the $\sim s^{\epsilon}$ rise of $\sigma_T$,
so that if $\sigma_{SD}$ is normalized to the 
experimental value at $\sqrt s\approx 22$ GeV the Regge prediction 
would exceed $\sigma_T$ at $\sqrt s\approx$ 2 TeV violating 
unitarity~\cite{RR}. The violation of unitarity is averted by renormalizing 
the ``Pomeron flux"~\cite{RR}
to unity, which is equivalent to renormalizing the 
rapidity gap probability of Eq.~(\ref{sddd}). This prescription leads 
to excellent agreement with data. It is interesting that the 
renormalization procedure, 
which violates Regge factorization, results in an energy independent 
$d\sigma_{SD}/dM^2$ cross section, which represents a scaling 
behaviour; moreover, it can also be applied to 
double-diffraction~\cite{KGgap}. In the present paper, these results are 
 obtained in 
a parton model approach, which has the added advantage that it can be 
generalized to describe processes with multiple rapidity 
gaps~\cite{snowmass}.
The most interesting conclusion from this approach 
is that in multigap events the 
renormalization factor depends only on the $s$-value of the collision,
so that the same suppression factor, otherwise known as ``gap survival 
probability", applies to a process regardless of the number of gaps 
it contains. The ratio of rates of two-gap to one-gap production 
in a diffractive process is predicted  
to be equal to the ratio of the $\pom\pom\pom$ to $\pom$-$p$
couplings, neglecting phase space edge effects. Thus, this factor is
interpreted as being the ``price to pay" for the color-matching required
to form a color-singlet with vacuum quantum numbers. The parton model 
approach can also be applied to hard diffraction 
(not discussed here for lack of 
space) and explains the breakdown of diffractive QCD factorization 
between HERA and the Tevatron~\cite{newapproach}.


\begin{thebibliography}{99}
\bibitem{snowmass}K. Goulianos, {\em The Nuts and Bolts of Diffraction},
arXiv:hep-ph/0110240, Presented at ``Snowmass2001, the future of
particle physics", Snowmass, CO, USA, July 2001.  
\bibitem{Regge}P.D.B. Collins, {\em An Introduction to Regge
Theory and High Energy Physics} (Cambridge University  Press, Cambridge 1977).
\bibitem{Feynman}R. Feynman, Phys. Rev. Lett. {\bf 23}, 1415 (1969).
\bibitem{dd}T. Affolder {\em et al.}, CDF Collaboration,  
Phys. Rev. Lett. {\bf 87}, 141802 (2001).
\bibitem{RR}K. Goulianos, Phys. Lett. {\bf B358}, 379 (1995).
\bibitem{KGgap} K. Goulianos, arXiv:hep-ph/9806384.
\bibitem{CMG}R.J.M. Covolan, J. Montanha and K. Goulianos, 
Phys. Lett. {\bf B389}, 176 (1996).
\bibitem{GM}K. Goulianos and J. Montanha, 
Phys. Rev. {\bf D50}, 114017 (1999).
\bibitem{Levin}See  E. Levin, {\em An Introduction to Pomerons},
arXiv:hep-ph/9808486, 
Presented at LAFEX International School on High-Energy Physics 
(LISHEP 98), session B: Advanced School in HEP,
Rio de Janeiro, Brazil, 16-20 Feb 1998. 
Published in ``Rio de Janeiro 1998, High energy physics", 261-336. 
\bibitem{jj1800}T. Affolder {\em et al.}, CDF Collaboration, 
Phys. Rev. Lett. {\bf 84}, 5083 (2000).
\bibitem{jj630}D. Acosta {\em et al.},  CDF Collaboration, 
accepted by Phys. Rev. Letters; T.~Affolder {\em et al.}, arXiv:hep-ex/0109025.
\bibitem{W}F. Abe {\em et al.}, CDF Collaboration, 
Phys. Rev. Lett. {\bf 78}, 2698 (1997).
\bibitem{jj}F. Abe {\em et al.}, CDF Collaboration, 
Phys. Rev. Lett. {\bf 79}, 2636 (1997).
\bibitem{b}T. Affolder {\em et al.},  CDF Collaboration,  
Phys. Rev. Lett. {\bf 84}, 232 (2000).
\bibitem{jpsi}T. Affolder {\em et al.},  CDF Collaboration, 
Phys. Rev. Lett. {\bf 87}, 241802 (2001); 
arXiv:hep-ex/0107071.
\bibitem{jjhera}C. Adloff {\em et al.}, H1 Collaboration, 
arXiv:hep-ex/0012151.
\bibitem{DPE}T. Affolder {\em et al.}, CDF Collaboration,  
Phys. Rev. Lett. {\bf 85}, 4217 (2000).
\bibitem{newapproach}K. Goulianos, J. Phys. {\bf G26}, 716 (2000).
\end{thebibliography}
\end{document}